\documentstyle[12pt]{article}

\def\Tr{\mathop{\rm Tr}\nolimits}

\newcommand{\VEV}[1]{\left\langle #1 \right\rangle}
\newcommand{\BigVEV}[1]{\Bigl\langle #1 \Bigr\rangle}

\newcommand{\hs}{\hspace*}

\newcommand{\nn}{\nonumber}

\begin{document}

\begin{titlepage}

\begin{flushright}
UT-842\\
YITP-99-21\\
\end{flushright}

\baselineskip 0.7cm

\vskip 2cm
\begin{center}
{\large \bf  Grand-Unification Scale Generation\\
             through the Anomalous U(1) Breaking}
\vskip 1.2cm

Izawa~K.-I.${}^{1,2}$,
Kiichi~Kurosawa${}^{1,3}$,
Yasunori~Nomura${}^1$,
and
T.~Yanagida${}^{1,2}$

\vskip 0.4cm

${}^{1}${\it Department of Physics, 
               University of Tokyo,\\
               Tokyo 113-0033, Japan}\\
${}^{2}${\it Research Center for the Early Universe,
               University of Tokyo,\\
               Tokyo 113-0033, Japan}\\
${}^{3}${\it Yukawa Institute for Theoretical Physics, 
               Kyoto University,\\
               Kyoto 606-8201, Japan}\\

\end{center}
\vskip 1.5cm
\abstract{
We discuss the anomalous U$(1)$ gauge symmetry as a mechanism 
of generating the grand-unification scale. 
We conclude that unification to a simple group
cannot be realized unless some parameters are ``tuned'',
and that models with product gauge groups are preferred.
We consider the ``$R$-invariant natural unification''
model with gauge groups SU$(5)_{GUT}\times$U$(3)_H$.
In this model the doublet-triplet splitting problem is solved and 
the unwanted GUT relation $m_s = m_\mu$ is avoided maintaining 
$m_b = m_\tau$.
Moreover, $R$-invariance suppresses the dangerous proton decays
induced by dimension four and five operators.
}
\end{titlepage}

\baselineskip 0.7cm

\section{Introduction}
Three gauge coupling constants of the standard-model gauge interactions
determined at the weak scale $\mu \simeq m_Z$ strongly suggest
supersymmetric (SUSY) grand unification (GUT) of all gauge groups 
SU$(3)_C$, SU$(2)_L$ and U$(1)_Y$ at a very high energy scale, $M_{GUT}
\simeq 2\times10^{16}$GeV. If the gravitational scale $M_* \simeq
2.4\times 10^{18}$GeV is the fundamental cut-off scale of 
field-theory description of nature, it implies the presence of a mild
hierarchy $M_{GUT}/M_* \simeq 10^{-2}$. 
It is very natural to consider such a hierarchy is a low-energy
manifestation of more fundamental theory. 

There is a mechanism which generates such a mild hierarchy. 
It is an anomalous U$(1)_A$ gauge symmetry whose anomalies are cancelled 
by Green-Schwarz mechanism~\cite{GS}. The anomalous U$(1)_A$ gauge
symmetry often appears in low energy effective theories of string
theories~\cite{anom}. Since the sum of U$(1)_A$ charges is non-zero, $\Tr
Q_A \neq 0$, in the anomalous U$(1)_A$ gauge symmetry, Fayet-Iliopoulos
(FI) term is induced at one-loop level~\cite{FI}. In pertubative
heterotic string theories it is given by
\begin{eqnarray}
{\cal L_{\rm FI}}= g_A^2\frac{\Tr Q_A}{192\pi^2}  M_*^2 D_A 
     \equiv  \xi^2  D_A.
\end{eqnarray} 
If all fields have no VEVs, D-term is non-zero so that SUSY is broken.
Instead, one usually assumes that some field $\phi$ has VEV 
to restore SUSY and define the normalization of U$(1)_A$ charge 
so that $\phi$ has a U$(1)_A$ charge $-1$.
In this normalization $\VEV{\phi}=\xi$ and
a mild hierarchy is generated as   
$\VEV{\phi}/M_* \equiv \epsilon \sim 
{\cal O}(10^{-1})$.

In this paper, we discuss the anomalous U$(1)_A$ gauge symmetry as 
a mechanism of generating the mild hierarchy $M_{GUT}/M_*$.\footnote{
There are several attempts to generate the hierarchy $M_{GUT}/M_*$
\cite{GUTscale}.} 
We first show that the
standard SUSY GUT based on a simple group cannot be realized unless some
parameters are ``tuned'' and that 
models with product gauge groups are preferred instead.
Here, we construct the ``$R$-invariant natural
unification'' model~\cite{R_inv} based on gauge groups
SU$(5)_{GUT}\times$U$(3)_H$, in which the symmetry breaking scale 
$\simeq 2\times 10^{16}$GeV is induced through the anomalous U$(1)_A$
breaking. We also show that this model has various phenomenologically
desirable features that the doublet-triplet splitting problem is solved,
proton decays induced by dimension four and five
operators~\cite{P_decay,dim_five} are suppressed by
$R$-symmetry and $m_s \neq m_\mu$ without violating the GUT relation,
$m_b = m_\tau$.  

\section{Standard SUSY GUT models}
\label{sec:2}
In this section, we show that it is difficult to generate the GUT scale
through anomalous U$(1)_A$ breaking in the standard SUSY SU$(5)$ GUT
models.
In order to break SU$(5)$ down to
SU$(3)_C\times$SU$(2)_L\times$U$(1)_Y$,
a Higgs field $\Sigma$ in adjoint representation 
must have an appropriate vacuum expectation value (VEV), 
$\VEV{\Sigma}/M_* \sim {\cal O}(10^{-2})$.
Since there is no scale but
the breaking scale of the anomalous U$(1)_A$ gauge
symmetry,
VEV of a field with a negative U$(1)_A$ charge $-q$ is controlled by its
U$(1)_A$ charge and is of order $\epsilon^q M_*$ in general.   
That is, the field $\Sigma$ is required to 
have a negative U$(1)_A$ charge $-2$.

Now, we give a simple example for generating the GUT scale.
Since $\Sigma$ has U$(1)_A$ charge $-2$, we have to 
introduce another field $\Sigma'$ in adjoint representation
with a positive U$(1)_A$ charge $4$
to have a nontrivial superpotential.
Consider the following superpotential,\footnote
{
The last term is necessary to give masses for 
$(\bf 3,2^\star)$ and $(\bf 3^\star,2)$ 
components in $\Sigma'$,
since $(\bf 3,2^\star)$ and $(\bf 3^\star,2)$ in $\Sigma$ 
are absorbed into broken gauge bosons.
}
\begin{eqnarray}
 W= \kappa \Sigma' 
    \left\{ \left(\frac{\phi}{M_*}\right)^2 M_* \Sigma + \kappa' 
       \Sigma^2\right\}
    + \kappa'' \left(\frac{\phi}{M_*}\right)^8  M_* {\Sigma'}^2,
\label{standard}
\end{eqnarray}
where parameters, $\kappa$'s, are assumed to be of order unity.
There is a desirable SUSY vacuum which breaks SU$(5)$ down to 
SU$(3)_C\times$SU$(2)_L\times$U$(1)_Y$,
\begin{eqnarray}
  \VEV{\Sigma}/M_* \simeq (\epsilon^2/\kappa'){\rm diag}(2,2,2,-3,-3).
\end{eqnarray}
The generated GUT scale is indeed 
${\cal O}(\epsilon^2 M_*)$ and the masses for broken gauge bosons,
$M_V$, and those for $(\bf 8,1)$, $(\bf 1,3)$ components 
in $\Sigma$ and $\Sigma'$,
$M_{\Sigma}$, are of order $\epsilon^2 M_*$,
while masses for $(\bf 3,2^\star)$ and $(\bf 3^\star,2)$ components
in $\Sigma'$, $M_X$, of order $\epsilon^{8} M_* \ll M_{GUT}$.
However, the disparity of the masses for various fields 
which belong to the same representation under SU$(5)$ 
disturbs the unification of gauge coupling constants.
We show it explicitly by considering the following combination of
standard-model gauge coupling constants.
From renormalization group equations (RGEs) at one-loop level, 
we get~\cite{HMY}
\begin{eqnarray}
\left(\frac{5}{\alpha_1}-\frac{3}{\alpha_2}-\frac{2}{\alpha_3}
\right)(m_Z)
&=& \frac{8}{2\pi} \ln \left(\frac{m_{SUSY}}{m_Z}\right)
  +\frac{12}{2\pi}\ln \left(\frac{M_V^2 M^2_\Sigma}{m_Z^3 M_X}\right).
\end{eqnarray}
The observed values for standard-model gauge couplings constrain 
the combination of masses in the last term as 
\begin{eqnarray}
\left(\frac{M_V^2 M_\Sigma^2}{M_X}\right)^{1/3} 
\simeq M_{GUT} \simeq 2\times 10^{16}{\rm GeV}.
\label{mass_relation}
\end{eqnarray}
On the other hand, $(M_V^2 M_\Sigma^2/M_X)^{1/3}$ is almost equal to 
$M_*$ in the present model, since factors $\epsilon$ in $M_V$, 
$M_\Sigma$ and $M_X$ are cancelled out. 
It implies that the standard-model gauge coupling constants do not unify 
at the high-energy scale due to the mass splitting among various fields
in the SU(5) multiplet $\Sigma'$.
We show, in Appendix, that such a situation cannot be avoided by adding
fields in  various representations.
One possibility for avoiding it is to ``tune'' some parameters. 
For example, if we assume that
$\kappa$ is not of order unity
but ${\cal O}(10^{-3})$ in the superpotential Eq.~(\ref{standard}), 
$M_\Sigma \simeq 10^{-3} \epsilon^2 M_*$ and the combination 
$(M_V^2 M_\Sigma^2/M_X)^{1/3}$ is $10^{-2} M_* \simeq 10^{16}$GeV. 

In this paper we pursue another possibility in which no ``tuning'' is
required. 
It implies that we discard a ``simple'' idea that the standard-model
gauge groups unify to one simple gauge symmetry.
That is, we need GUT models with product gauge groups.
As such models, we consider ``$R$-invariant natural unification'' 
model with a gauge group SU$(5)\times$U$(3)_H$~\cite{R_inv},
since the anomalous U$(1)_A$ gauge symmetry is naturally 
embedded in this model.
Moreover, it has various desirable features in view of phenomenology.

\section{$R$-invariant unification model}
Let us first discuss briefly the ``$R$-invariant natural unification''
model which is based on a SUSY SU$(5)_{GUT}\times$U$(3)_H$ 
gauge theory with an unbroken $R$ symmetry~\cite{R_inv}. 
The SU$(5)_{GUT}$ is the usual GUT gauge 
group and its coupling constant is in a perturbative regime, while the
U$(3)_H$ is a hypercolor gauge group whose coupling is strong at
the GUT scale. Here, GUT means unification of all gauge groups,
SU$(3)_C$, SU$(2)_L$ and U$(1)_Y$, in the standard model. The quarks
and leptons obey the usual transformation law under the GUT group
SU$(5)_{GUT}$ and they are all singlets of the hypercolor group
U$(3)_H$. A pair of Higgs multiplets $H_r$ and $\bar H^r$
($r=1,\cdots,5$) transform as ${\bf 5}+{\bf 5^\star}$ under the 
SU$(5)_{GUT}$ and as singlets under the U$(3)_H$.

All the matter multiplets introduced so far are the same as in the
minimal SUSY SU$(5)$ model. We now introduce six pairs of
hyperquarks $Q^\rho_\alpha$ and $\bar Q^\alpha_\rho$ 
($\alpha=1,\cdots,3$; $\rho=1,\cdots,6$) which transform as ${\bf 3}$
and ${\bf 3^\star}$ under the 
hypercolor SU$(3)_H$ and have U$(1)_H$ charges $1$ and $-1$,
respectively (SU$(3)_H\times$U$(1)_H\equiv$U$(3)_H$). 
The first five pairs of $Q_\alpha^r$ and $\bar Q^\alpha_r$
($r=1,\cdots,5$) belong to ${\bf 5^\star}$ and ${\bf 5}$ of SU$(5)_{GUT}$
and the last pair of $Q_\alpha^6$ and $\bar Q^\alpha_6$ are singlets of
SU$(5)_{GUT}$.  To cause a breaking of the total gauge group 
SU$(5)_{GUT}\times$U$(3)_H$ down to the standard-model gauge groups,
SU$(3)_C\times$SU$(2)_L\times$U$(1)_Y$, we introduce another chiral 
multiplet $X^\alpha_\beta$ coupled to $Q_\alpha^r$ and $\bar Q^\alpha_r$, 
which is an adjoint representation of the SU$(3)_H$. 
Since $Q_\alpha^r$
and $\bar Q^\alpha_r$ are supposed to have VEV
of order of the GUT scale, they must be trivial representations
of the $R$ symmetry U$(1)_R$ \cite{R_inv}, and hence the $X^\alpha_\beta$ carry 
$R$-charge two. $R$ charges for all matter multiplets besides quark and
lepton multiplets are given in Table~\ref{charge}.

To suppress unwanted nonrenormalizable interactions in superpotential we 
further impose an axial U$(1)_A$ symmetry, under which the hyperquarks
$Q_\alpha^r$ and $\bar Q^\alpha_r$ and the adjoint $X^\alpha_\beta$
transform as
\begin{eqnarray}
   Q_\alpha^r \to e^{i\theta} Q_\alpha^r,
\hs{5mm}
  \bar Q^\alpha_r \to e^{i\theta} \bar Q^\alpha_r,
\hs{5mm}
  X^\alpha_\beta \to e^{-2i\theta} X^\alpha_\beta.
\end{eqnarray} 
U$(1)_A$ charges for all matter multiplets besides quark and lepton
multiplets are also given in
Table~\ref{charge}. 
\begin{table}[h]
\begin{eqnarray}
\begin{array}{|c|cc|cc|c|cc||cc|}
\hline
 & Q_\alpha^r & \bar Q^\alpha_r & Q^6_\alpha & \bar Q^\alpha_6 & 
 X^\alpha_\beta & H_r & \bar H^r & \phi & \chi\\
\hline
{\rm U}(1)_R & 0 & 0 & 2 & 2 & 2 & 0 & 0 & 0 & 2\\
{\rm U}(1)_A & -2 & -2 & 2 & 2 & 4 & 0 & 0 & -1 & 4\\
\hline
\end{array}
\nn
\end{eqnarray}
\caption{Charge assignments in the Higgs sector.}
\label{charge}
\end{table}
\\
Then, we have a superpotential\footnote
{
The imposition of U$(1)_A$ and U$(1)_R$ alone allows nonrenormalizable
interactions which include arbitrary powers of $H_r \bar H^r /M_*^2$.
}
\begin{eqnarray}
W=\lambda Q_\alpha^r \bar Q^\beta_r X^\alpha_\beta
  +h Q^r_\alpha \bar Q^\alpha_6 H_r
  +\bar h Q_\alpha^6 \bar Q^\alpha_r \bar H^r.
\label{Rinvariant}
\end{eqnarray}
Notice that the $R$ charges for $H_r$ and $\bar H^r$ are vanishing and
those for $Q_\alpha^6$ and $\bar Q^\alpha_6$ are two.
As shown in Ref.~\cite{R_inv} we have the desirable vacua;
\begin{eqnarray}
\BigVEV{X^\alpha_\beta}=\BigVEV{Q_\alpha^6}=\BigVEV{\bar Q^\alpha_6}=0,
\hs{3mm}
\BigVEV{H_r}=\BigVEV{\bar H^r}=0,
\hs{3mm}
\BigVEV{Q_\alpha^r} = v \delta_\alpha^r,
\hs{3mm}
\BigVEV{\bar Q^\alpha_r} = v \delta^\alpha_r.
\label{vacuum}
\end{eqnarray}

For $v\neq 0$, the total gauge group, SU$(5)_{GUT}\times$U$(3)_H$, is 
broken down to SU$(3)_C$ $\times$SU$(2)_L\times$U$(1)_Y$ and hence the $v$
corresponds to the GUT scale. In these vacua, the color SU$(3)_C$ is 
an unbroken linear combination of an SU$(3)$ subgroup of the 
SU$(5)_{GUT}$ and the hypercolor SU$(3)_H$, and the hypercharge U$(1)_Y$ 
is that of a U$(1)$ subgroup of the SU$(5)_{GUT}$ and the strong
U$(1)_H$. Thus, the gauge coupling constants $\alpha_3$, $\alpha_2$ and
$\alpha_1$ of SU$(3)_C\times$SU$(2)_L\times$U$(1)_Y$ are given by  
\begin{eqnarray}
\alpha_3 \simeq \frac{\alpha_{\rm GUT}}
                    {1+\alpha_{\rm GUT}/\alpha_{3H}},
\hs{3mm}
\alpha_2 = \alpha_{\rm GUT},
\hs{3mm}
\alpha_1 \simeq \frac{\alpha_{\rm GUT}}
                     {1+\frac{1}{15}\alpha_{\rm GUT}/\alpha_{1H}},
\end{eqnarray}
where $\alpha_{3H}$ and $\alpha_{1H}$ are gauge coupling constants for
the hypercolor SU$(3)_H$ and U$(1)_H$, respectively. We see that the
unification of three gauge coupling constants, $\alpha_3$, $\alpha_2$
and $\alpha_1$, is practically achieved in a strong coupling regime of
the hypercolor gauge interactions, that is, $\alpha_{3H}$ and
$\alpha_{1H} \gg {\cal O}(1)$.

Interesting is that the color triplets $H_a$ and $\bar H^a$
($a=1,\cdots,3$) acquire masses of order $v$ together with the sixth
hyperquarks $\bar Q^\alpha_6$ and $Q^6_\alpha$ in the vacua
Eq.~(\ref{vacuum}), while the weak doublets $H_l$ and $\bar H^l$
($l=4,5$) remain massless since there are no partners for them to form
$R$-invariant masses. The masslessness for these doublets is guaranteed
by the unbroken U$(1)_R$ symmetry.\footnote
{
The U$(1)_R$ symmetry is broken down to a discrete subgroup $Z_{4R}$ by
the hypercolor SU$(3)_H$ anomaly. However, the unbroken subgroup
$Z_{4R}$ is sufficient to keep the Higgs doublets massless.
}

So far, the GUT scale $v$ is undetermined because of the presence of a
flat direction in the present vacua. We identify the axial U$(1)_A$ with
the anomalous U$(1)_A$ gauge symmetry and
show that the GUT scale $v$ is determined by the breaking scale of the
anomalous U$(1)_A$.
To obtain the desirable value for $v\simeq 2\times10^{16}$GeV
we introduce a singlet $\chi$ with a U$(1)_A$ charge $+4$
and the following superpotential
\begin{eqnarray}
W = k Q_\alpha^r \bar Q^\alpha_r \chi + \frac{k'}{M_*^2} \phi^4 \chi.
\end{eqnarray}
Here, we have assumed $R$ charges for $\chi$ and $\phi$ are two and zero,
respectively. Then, we get the GUT scale with a correct magnitude
\begin{eqnarray}
  v \simeq \sqrt{\frac{k'}{k}} \frac{\VEV{\phi}^2}{M_*} 
    \sim {\cal O}(10^{16}{\rm GeV}),
\end{eqnarray}
for $k,k' \sim  {\cal O}(1)$.

\section{Quark and lepton mass matrices}
Let us turn to discuss quark and lepton mass matrices. It is very
attractive to use the above $\phi$ field generating hierarchies in quark
and lepton mass matrices~\cite{FN,Yukawa}. We tentatively 
assume U$(1)_A$ charges for quark and lepton multiplets, ${\bf
5}^\star_i$ and ${\bf 10}_i$ ($i=1,\cdots,3$) as shown in
Table~\ref{FN_charge}~\cite{large}
where U$(1)_R$ charges are also given. 
There are two cases for the assignment of $U(1)_A$ charges 
corresponding that $\tan\beta$ is small or large,
where $\tan\beta$ is the ratio of the VEVs of the Higgs
($\tan\beta=\VEV{H}/\langle\bar H\rangle$).
That is, $\tau=1$ for small $\tan\beta \sim {\cal O}(1)$ and 
$\tau= 0$ for large $\tan\beta\sim {\cal O}(1/\epsilon)$.
\begin{table}[h]
\begin{eqnarray}
\begin{array}{|c|ccc|ccc||cc|}
\hline
 & {\bf 10}_1 & {\bf 10}_2 & {\bf 10}_3 & {\bf 5}^\star_1 & {\bf 5}^\star_2 & {\bf 5}^\star_3 
& H'({\bf 45}) & \bar H'({\bf 45^\star})\\
\hline
{\rm U}(1)_R & 1 & 1 & 1 & 1 & 1 & 1 & 0 & 2\\
{\rm U}(1)_A & 2 & 1 & 0 & \tau+1 & \tau & \tau & -\tau-1 & 4\\
\hline
\end{array}
\nn
\end{eqnarray}
\caption{Charge assignments for the matter fields.}
\label{FN_charge}
\end{table}

Because of the U$(1)_A$ invariance and $\VEV{\phi}/M_* = \epsilon$, 
we obtain the Yukawa matrix for up-type quarks as
\begin{eqnarray}
 \widehat{\lambda}_u \simeq
\left(
\begin{array}{ccc}
\epsilon^4 & \epsilon^3  & \epsilon^2 \\
\epsilon^3 & \epsilon^2  & \epsilon   \\
\epsilon^2 & \epsilon    & 1
\end{array}
\right),
\end{eqnarray}
and those for down-type quarks and charged leptons as
\begin{eqnarray}
 \widehat{\lambda}_d =\widehat{\lambda}_l \simeq
\epsilon^{\tau} \left(
\begin{array}{ccc}
\epsilon^3 & \epsilon^2  & \epsilon^2\\
\epsilon^2 & \epsilon    & \epsilon\\
\epsilon   & 1           & 1
\end{array}
\right).
\label{wrong_relation}
\end{eqnarray}
Each element in the matrices has an undetermined coefficient of 
order unity. If $\epsilon \simeq 1/20$, 
the above mass matrices are roughly consistent with observations except
for the unwanted GUT relations
\begin{eqnarray}
  m_s=m_\mu \;\;\mbox{and}\;\; m_d=m_e.
\end{eqnarray}
It is only the GUT condensation,
$\BigVEV{Q^r_\alpha}=v \delta^r_\alpha$ and 
$\BigVEV{\bar Q^\alpha_r}=v
\delta^\alpha_r$, that can break these unwanted GUT relations. 
In the above model, however, a possible operator is only
\begin{eqnarray}
 \frac{1}{M_*^2} {\bf 5}^\star_1 {\bf 10}_1 \bar H 
 \VEV{Q_\alpha \bar Q^\alpha}
\end{eqnarray}
for $\tau=1$ and nothing for $\tau=0$.
Thus, it cannot lead to sufficient contributions to the 
quark and lepton mass matrices. 
A way to solve this problem is to introduce another pair of Higgs
multiplets $H'$ and $\bar H'$ transforming as ${\bf 45}$ and ${\bf
45^\star}$ under the SU$(5)_{\rm GUT}$.\footnote{
Another way to solve the problem is to lower the cut-off scale,
$M_*$, and change the U$(1)_A$ charge assignments in 
Table~\ref{charge}.
}
We assign U$(1)_A$ charge for ${\bf
45}$ so that the following Yukawa coupling is possible,
\begin{eqnarray}
   {\bf 5}^\star_2 {\bf 10}_2 H'({\bf 45}).
\end{eqnarray}
On the other hand, since the existence of 
${\bf 5}^\star_3 {\bf 10}_3 H'({\bf 45})$ would
break the successful GUT relation $m_b=m_\tau$,
U$(1)_A$ and U$(1)_R$ charges for $H'({\bf 45})$ 
are determined as 
in Table~\ref{FN_charge}.

Now, massless Higgs doublet $\bar H_f$ is a linear combination of
doublets, $\bar H_f({\bf 5}^\star)$ in $\bar H({\bf 5}^\star)$ and 
$H'_f({\bf 45})$ in $H'({\bf 45})$. 
Mixing angle $\theta$ is
determined by the following superpotential
\begin{eqnarray}
W= H'({\bf 45}) \bar H'({\bf 45}^\star) \frac{\phi^2}{M_*} 
    \;(\;{\rm or}\;\frac{\phi^3}{M_*^2}\;)\;
  + \frac{1}{M_*} \bar H({\bf 5}^\star) \bar H'({\bf 45}^\star) 
    \VEV{Q_\alpha \bar Q^\alpha},
\end{eqnarray}
for $\tau =1$ (or $\tau =0$).
Here, we have taken U$(1)_A$ and U$(1)_R$ charges for 
$\bar H'({\bf 45^\star})$ as in Table~\ref{FN_charge}.
We get $\theta \simeq \VEV{\phi}^2/M_*^2$ ( or $\VEV{\phi}/M_*$ ) 
where $\theta$ is defined as
\begin{eqnarray}
\bar H_f=\bar H_f({\bf 5}^\star)\cos\theta +H'_f({\bf 45}) \sin\theta.
\end{eqnarray}
This gives 
\begin{eqnarray}
     \VEV{H'_f({\bf 45})} & \simeq & \epsilon^2 \;({\rm or}\; \epsilon)
          \VEV{\bar H_f},\\
     \VEV{\bar H_f({\bf 5}^\star)}& \simeq & \VEV{\bar H_f},
\end{eqnarray}
which leads to a required magnitude of the violation of the unwanted 
GUT relations, $m_s = m_\mu$ and $m_d = m_e$, as follows:
\begin{eqnarray}
\delta \widehat{\lambda}_d \simeq
 (-2)\times\epsilon^{\tau}\left(
\begin{array}{ccc}
\epsilon^3 & \epsilon^2  & \epsilon^2\\
\epsilon^2 & \epsilon    & \epsilon  \\
\epsilon   & 0           & 0
\end{array}
\right), 
\hs{1cm}
\delta \widehat{\lambda}_l \simeq
3\times\epsilon^{\tau}\left(
\begin{array}{ccc}
\epsilon^3 & \epsilon^2  & \epsilon^2\\
\epsilon^2 & \epsilon    & \epsilon  \\
\epsilon   & 0           & 0
\end{array}
\right).
\end{eqnarray}
Note that the contribution from the GUT condensation for down-type
quarks has an opposite sign to that for charged leptons. It shows that
we can get the relation $m_s < m_\mu$ when 
the signs of $(2,2)$ elements in $\widehat{\lambda}_d$ and 
$\delta\widehat{\lambda}_d$ are opposite.
If the GUT condensation is not taken into account,
Cabbibo angle is given by $\epsilon$, which is smaller than the
observed value $0.22$. However, due to the cancellation 
in (2,2) element in the mass matrix for down-type quarks,
we can get the desirable value.

\section{Anomaly cancellations}

In this section, we discuss the cancellation of the U$(1)_A$ mixed
anomalies, $C_G$, with the gauge groups $G$ 
($G = $SU$(5)_{GUT}$, SU$(3)_H$ and U$(1)_H$).
In the above model, we obtain $C_G$ as
\begin{eqnarray}
\begin{array}{lcc}
 C_{{\rm SU}(5)_{GUT}}&= & 35-\frac{21}{2}\tau,  \\
 C_{{\rm SU}(3)_H}    &= & 4,  \\
 C_{{\rm U}(1)_H}     &= & -48.
\end{array}
\label{anomaly}
\end{eqnarray}
We find that U$(1)_A$ has the negative anomaly coefficient 
$C_{{\rm U}(1)_H}$, which may not be expected in usual heterotic 
string theories.\footnote
{
Apparently, it seems possible that we add U$(1)_H$-charged fields with
positive U$(1)_A$ charges so that the anomaly coefficient 
$C_{{\rm U}(1)_H}$ becomes positive.
In that case, however, the gauge coupling constant for U$(1)_H$ blows up 
below $M_*$.
}
In the heterotic string case there is the {\it universal} relation 
to the mixed anomaly coefficients as follows: 
\begin{eqnarray}
   \frac{C_G}{k_G} = {\rm const.},  
\end{eqnarray} 
where $k_G$ is the Kac-Moody level, since only the dilaton field $S$
plays a role in anomaly cancellations or, in other words, there is 
only one antisymmetric tensor field, $B_{\mu\nu}$ 
(Green-Schwarz mechanism)~\cite{GS}.

However, we can consider the case where various moduli fields other than 
the dilaton cancel these anomalies 
(``generalized'' Green-Schwarz mechanism).
Indeed, such a situation is realized in Type I and Type IIB string
theories with orientifold compactifications~\cite{Type12}.
The anomalous U(1) gauge symmetries in these theories have been studied 
recently and it has been revealed that several anomalous U$(1)$ gauge
symmetries exist corresponding to the existence of several
antisymmetric tensors in twisted sectors.

The ``generalized'' Green-Schwarz mechanism is illustrated as follows.
In this mechanism there are many moduli fields $M_k$ which are coupled
to field strength superfields as
\begin{eqnarray}
{\cal L}_{gauge} 
  = \int d^2\theta \left( k_G S + \sum_k c_G^k M_k \right) 
    W_G^\alpha W_{G\alpha},
\end{eqnarray}
and these moduli fields transform under the U$(1)_A$ transformation, 
$V_A \rightarrow V_A + \frac i2 (\Lambda - \Lambda^{\dagger})$,
as 
\begin{eqnarray}
  M_k \rightarrow M_k + i\delta_k \Lambda,
\end{eqnarray}
where $c_G^k$ and $\delta_k$ are model-dependent constants.
This gives the {\it nonuniversal} relation to the 
mixed anomaly coefficients
\begin{eqnarray}
   C_G = 4\pi^2 \sum_k c_G^k \delta_k.
\end{eqnarray} 
It shows that the three anomalies in Eqs.~(\ref{anomaly}) can be
cancelled by the non-linear transformation of the moduli fields $M_k$.

The origin of FI term is also different from that in the case of 
perturbative heterotic string theories.
FI term is not generated at one-loop level but is given 
by the VEVs of moduli fields $\VEV{M_k}$ at tree level as
\begin{eqnarray}
{\cal L}_{FI}
  \simeq -\sum_k \delta_k \VEV{M_k} M_*^2 D_A \equiv \xi^2 D_A.    
\end{eqnarray}
Although the VEVs of the moduli fields $\VEV{M_k}$ are undetermined 
in the present framework, it seems natural to consider that the
resulting FI term is one order smaller than $M_*$, 
i.e. $\epsilon = \xi/M_* \sim 10^{-1}$.
Then, the conclusions in the previous sections are not affected.

\section{Summary and Discussions}
We have found that the idea that the GUT scale is generated through the 
anomalous U$(1)_A$ breaking is compatible with the ``$R$-invariant
natural unification'' model with gauge groups 
SU$(5)_{GUT}\times$U$(3)_H$. 
The Higgs field which breaks the GUT group has an 
anomalous U$(1)_A$ charge and the unwanted GUT relations, 
$m_s = m_\mu$ and $m_d = m_e$,
can be avoided in a simple manner without violating the successful
GUT relation, $m_b=m_\tau$.
Also, the doublet-triplet splitting problem is solved naturally
in the present model. 

As for proton decay, the vacuum in this model preserves the U$(1)_R$
invariance and $R$-parity is included in the U$(1)_R$ symmetry.
Therefore, the dimension four and five operators contributing to
the proton decay~\cite{P_decay,dim_five} are suppressed. 
Instead, the process induced by the dimension six operators
is dominant.

It is a crucial difference from the usual GUT models that 
the SU$(3)_C$ is an unbroken linear combination of an SU$(3)$ 
subgroup of the SU$(5)_{GUT}$ and the hypercolor SU$(3)_H$.
It predicts a smaller value of the strong coupling constant $\alpha_3$.
The current experimental values of $\alpha_3$~\cite{PDG}
are consistent with the present model~\cite{gaugino}.

We conclude this paper with a comment on neutrino masses~\cite{neutrino}.
Since neutrino masses are written by the effective operators, 
$({\bf 5}^\star H)^2$, they are independent of
U$(1)_A$ charges of right handed neutrinos and are given 
by,~\cite{large}  
\begin{eqnarray}
\widehat{m}_\nu \simeq
\epsilon^{2\tau}
\left(
\begin{array}{ccc}
\epsilon^2& \epsilon &\epsilon\\
\epsilon & 1 & 1\\
\epsilon & 1 & 1
\end{array}
\right) \frac{m^2_t}{M_*}.
\label{neutrino_mass}
\end{eqnarray}
This mass matrix implies 
the large angle between the second and third generations
and either small or large 
angle between the first and second ones. 
However, the mass of tau neutrino is 
of order of $10^{-5}$eV 
for $\tan\beta \sim {\cal O}(\epsilon^{-1})$ 
or $10^{-7}$eV for $\tan\beta \sim {\cal O}(1)$.
It is smaller by several orders than the scale indicated by
the atmospheric neutrino oscillation~\cite{SuperK}.
There are two ways to generate a desirable mass scale. One way 
is to introduce the U$(1)_{B-L}$ breaking scale as well as
the anomalous U$(1)_A$ breaking scale. The other is to 
extend the assignment of U$(1)_A$ charges given in Table~\ref{charge}
and Table~\ref{FN_charge}.
This can be done because the assignment of U$(1)_A$ charges 
is not uniquely determined by the following conditions:
$Q\bar Q$ has charge $-4$, the mixed anomaly U$(1)_H$-U$(1)_A^2$
vanishes, $H \bar H$ has charge $0$ in order to use the
Giudice-Masiero mechanism for $\mu$ and $B\mu$ terms~\cite{GM},
and the superpotential (\ref{Rinvariant}) is U$(1)_A$ invariant. 
The U$(1)_A$ charges for various fields are given in 
Table~\ref{neutrino}.
\begin{table}[h]
\begin{eqnarray}
\begin{array}{|c|cc|cc|c|cc||cc|}
\hline
 & \makebox[14mm]{$Q_\alpha^r$} & \makebox[14mm]{$\bar Q^\alpha_r$} & 
   \makebox[14mm]{$Q^6_\alpha$} & \makebox[14mm]{$\bar Q^\alpha_6$} & 
 \makebox[9.6mm]{$X^\alpha_\beta$} & 
 \makebox[10mm]{$H_r$} & \makebox[10mm]{$\bar H^r$} & 
 \makebox[9mm]{$\phi$} & \makebox[9mm]{$\chi$}\\
\hline
{\rm U}(1)_A & -2-p & -2+p & 2-5p & 2+5p & 4 & -4p & 4p & -1 & 4\\
\hline
\end{array}
\nn\\
\begin{array}{|c|ccc|ccc||cc|}
\hline
 & {\bf 10}_1 & {\bf 10}_2 & {\bf 10}_3 & {\bf 5}^\star_1 & {\bf 5}^\star_2 & {\bf 5}^\star_3 
& H'({\bf 45}) & \bar H'({\bf 45^\star})\\
\hline
{\rm U}(1)_A & 2+2p & 1+2p & 2p & \tau+1-6p & \tau-6p & \tau-6p & -\tau-1+4p & 4-4p\\
\hline
\end{array}
\nn
\end{eqnarray}
\caption{Assignment of U$(1)_A$ charges. $p$ is an arbitrary parameter.
Taking $p=0$ corresponds to the assignment given in Table~\ref{charge} and \ref{FN_charge}.}
\label{neutrino}
\end{table}

For this charge assignment
we are to replace the coefficient $\epsilon^{2\tau}$ with
$\epsilon^{-20p+2\tau}$ in Eq.~(\ref{neutrino_mass}). 
If we take $p=1/4$ and $\tau=1$, we can get 
desirable tau neutrino mass,
$m_{\nu_\tau} \simeq \epsilon^{-3} m_t^2/M_* \sim 0.1$eV. 
With this charge assignment, dimension four baryon-number-violating
operators are completely suppressed by an unbroken discrete gauge
symmetry.
It is also desirable that no other scales are required to understand the 
neutrino masses.

\section*{Acknowledgments}
Y.N. thanks the Japan Society for the Promotion of Science for financial 
support. This work is supported in part by the Grant-in-Aid, Priority
Area ``Supersymmetry and Unified Theory of Elementary Particles''(\#707). 

\appendix
\section*{Appendix}
In this appendix, we show by using the RGEs
that the unification to SU$(5)$
cannot be realized without ``tuning'' if all scales are
generated by the breaking of the anomalous U$(1)_A$ gauge symmetry.
We consider generic SU$(5)$ GUT models in which 
there are matter fields $\Psi_I$
with U$(1)_A$ charges $q_I$ 
in various representations {\boldmath $R_I$} of SU$(5)$
in addition to quark and lepton multiplets, 
${\bf 5^\star}_i$ and ${\bf 10}_i$, and the Higgs multiplets, 
$H({\bf 5})$ and $\bar H({\bf 5^\star})$.
The adjoint Higgs $\Sigma$ and $\Sigma'$ are included 
in $\Psi_I$. 
Owing to the breaking of GUT, 
the field in a representation {\boldmath $R_I$} 
is decomposed into fields $\psi_{I,i}$
in representations {\boldmath $r_{I,i}$} of
SU$(3)_C\times$SU$(2)_L\times$U$(1)_Y$, 
which have SUSY masses $M_{I,i}$.

From the RGEs at one-loop level, we get
\begin{eqnarray}
\left(
\begin{array}{c}
1/\alpha_3 \\ 
1/\alpha_2 \\ 
1/\alpha_1
\end{array}
\right)(m_Z)
&=&
\left(
\begin{array}{c}
1 \\ 
1 \\ 
1
\end{array}
\right)
\left\{\frac{1}{\alpha_5(M_*)}-\frac{3}{2\pi}
\ln\left(\frac{M_*}{m_Z}\right)\right\}
-\frac{1}{2\pi}
\left(
\begin{array}{c}
4 \\ 
25/6 \\ 
5/2
\end{array}
\right)
\ln \left(\frac{m_{SUSY}}{m_Z}\right)\nn\\
&&
+\frac{1}{2\pi}
\left(
\begin{array}{c}
4 \\ 
6 \\ 
10
\end{array}
\right)
\ln \left(\frac{M_V}{m_Z}\right)
-\frac{1}{2\pi} \times 2
\left(
\begin{array}{c}
1/2 \\ 
0 \\ 
1/5
\end{array}
\right)
\ln \left(\frac{M_{H_C}}{m_Z}\right)\nn\\
&&
-\frac{1}{2\pi}
{\sum_{(I,i)}}^\prime
\left(
\begin{array}{c}
T_3 (\mbox{\boldmath $r_{I,i}$}) \\ 
T_2 (\mbox{\boldmath $r_{I,i}$})\\ 
T_1 (\mbox{\boldmath $r_{I,i}$})
\end{array}
\right)
\ln \left(\frac{M_{I,i}}{m_Z}\right),
\label{RGE}
\end{eqnarray}
where $M_V$ is the mass of the broken gauge bosons 
and $M_{H_C}$ is that of the colored Higgs.
$T_n(\mbox{\boldmath $r$})$ ($n=1,2,3$) 
is the half of the Dynkin index for representation {\boldmath $r$}.
The primed sum in the last term
means that Goldstone modes for SU$(5)$ breaking, 
$({\bf 3,2^\star})_{-5/6}$ and $({\bf 3^\star,2})_{5/6}$, are not
included.
Now, we take the following linear combination 
of standard-model gauge coupling constants
in order to
drop the contributions from
fields in fundamental representations or its complex conjugates,
${\bf 5}$ and ${\bf 5^\star}$, including the colored Higgs, 
\begin{eqnarray}
\left(\frac{5}{\alpha_1}-\frac{3}{\alpha_2}-\frac{2}{\alpha_3}
\right)(m_Z)
\!\!\!&=&\!\!\! 
\frac{8}{2\pi} \ln \left(\frac{m_{SUSY}}{m_Z}\right)
  +\frac{24}{2\pi}\ln \left(\frac{M_V}{m_Z}\right)\nn\\
&&\!\!\!\!\!\!\!
+\frac{1}{2\pi} {\sum_{(I,i)}}^\prime 
\Bigl\{2T_3(\mbox{\boldmath $r_{I,i}$})
+3T_2(\mbox{\boldmath $r_{I,i}$})-5T_1(\mbox{\boldmath $r_{I,i}$})
\Bigr\}\ln
\left(\frac{M_{I,i}}{m_Z}\right).
\label{comb_RGE}
\end{eqnarray}
The sum in the last term can be rewritten as the sum of the
contributions from various representations {\boldmath $r$} 
of SU$(3)_C\times$SU$(2)_L\times$U$(1)_Y$,
\begin{eqnarray}
\frac{1}{2\pi}\sum_{\mbox{\boldmath $r$}}
\Bigl\{
2T_3(\mbox{\boldmath $r$})+3T_2(\mbox{\boldmath $r$})
-5T_1(\mbox{\boldmath $r$})
\Bigr\}
\ln\left[
\mathop{{\prod_{(I,i)}}^\prime}
   _{\mbox{\boldmath $r_{I,i}$}=\mbox{\boldmath $r$}} 
\left(\frac{M_{I,i}}{m_Z}\right)\right].
\end{eqnarray}
Recall that $M_{I,i}$ are eigenvalues for the mass matrices, 
$M^{\mbox{\boldmath $r$}}$, for representations {\boldmath $r$} 
which are determined when a specific superpotential is given.
Even if we do not specified a superpotential, however, we are 
able to see the structure of the mass matrices 
using their U$(1)_A$ charges as follows, 
\begin{eqnarray}
\mathop{{\prod_{(I,i)}}}
   _{\mbox{\boldmath $r_{I,i}$}=\mbox{\boldmath $r$}} 
M_{I,i} &=& \det
M^{\mbox{\boldmath $r$}}_{(I,i),(\bar J,\bar \jmath)}
\simeq \det \left[\epsilon^{q_I+q_{\bar J}} M_*\right]\nn\\ 
&\simeq&\left\{\mathop{{\prod_{(I,i)}}}
   _{\mbox{\boldmath $r_{I,i}$}=\mbox{\boldmath $r$}}
    \epsilon^{q_I} M_*^{1/2}
   \right\}
   \left\{
   \mathop{{\prod_{(\bar J,\bar \jmath)}}} 
   _{\mbox{\boldmath $r_{\bar J,\bar \jmath}$}=
              \mbox{\boldmath $r^\star$}} 
    \epsilon^{q_{\bar J}} M_*^{1/2}
   \right\}.
\label{mass}
\end{eqnarray}
Note that this equation is not valid for 
{\boldmath $r$}$=({\bf 3,2^\star})_{-5/6}$ 
and $({\bf 3^\star,2})_{5/6}$,
since the mass matrices for $({\bf 3,2^\star})_{-5/6}$ 
and $({\bf 3^\star,2})_{5/6}$ 
have a zero eigenvalue corresponding to the Goldstone mode.
Nevertheless, the relation similar to Eq.~(\ref{mass}) holds also 
for {\boldmath $r$}$=({\bf 3,2^\star})_{-5/6}$ including 
the mass of broken gauge boson, $M_V$.

Suppose that the SU$(5)$ multiplet $\Psi_X$ includes the Goldstone 
modes and has a negative anomalous U$(1)_A$ charge $q_X<0$.
Then, the mass for the broken gauge bosons is given by
\begin{eqnarray}
M_V \simeq g_5 \VEV{\Psi_X} \simeq \epsilon^{-q_X} M_*.
\end{eqnarray}
The second term in Eq.~(\ref{comb_RGE}) is rewritten as 
\begin{eqnarray}
\frac{24}{2\pi}
\ln\left(\frac{M_V}{m_Z}\right)
&\simeq& (-6)\times \frac{(-4)}{2\pi}\ln
\left(\frac{ \epsilon^{-q_X} M_*}{m_Z}\right)\\
&=&(-6) \times \frac{4}{2\pi}
\ln \left(\frac{\epsilon^{q_X} M_*^{1/2}}{m_Z^{1/2}}\right)
+\frac{12}{2\pi}\ln \left( \frac{M_*}{m_Z}\right)^3,
\label{mass_X}
\end{eqnarray}
and the first term in Eq.~(\ref{mass_X}) can be combined with 
the mass matrices for {\boldmath $r$}$=({\bf 3,2^\star})_{-5/6}$ 
and $({\bf 3^\star,2})_{5/6}$, giving the relation similar to 
Eq.~(\ref{mass}) as
\begin{eqnarray}
\left(\frac{\epsilon^{q_X} M_*^{1/2}}{m_Z^{1/2}}\right)
+\mathop{{\prod_{(I,i)}}^\prime}
   _{\mbox{\boldmath $r_{I,i}$}=\mbox{\boldmath $r$}} 
M_{I,i} 
= 
\left(\frac{\epsilon^{q_X} M_*^{1/2}}{m_Z^{1/2}}\right)
+{\det}^\prime
M^{\mbox{\boldmath $r$}}_{(I,i),(\bar J,\bar \jmath)}
\simeq \det \left[\epsilon^{q_I+q_{\bar J}} M_*\right].
\label{mass_X_V}
\end{eqnarray}
Using Eqs.~(\ref{mass}, \ref{mass_X_V}), the Eq.~(\ref{comb_RGE}) 
is expressed as
\begin{eqnarray}
\left(\frac{5}{\alpha_1}-\frac{3}{\alpha_2}-\frac{2}{\alpha_3}
\right)(m_Z)
\!\!\!&=&\!\!\! 
\frac{8}{2\pi} \ln \left(\frac{m_{SUSY}}{m_Z}\right)
 +\frac{12}{2\pi}\ln \left(\frac{M_*}{m_Z}\right)^3\nn\\
&&\hs{-3cm}
+\frac{2}{2\pi} \sum_I \ln\left(
\frac{\epsilon^{q_I} M_*^{1/2}}{m_Z^{1/2}}\right)
\left[\sum_i 
\Bigl\{2T_3(\mbox{\boldmath $r_{I,i}$})
+3T_2(\mbox{\boldmath $r_{I,i}$})-5T_1(\mbox{\boldmath $r_{I,i}$})
\Bigr\}\right].
\end{eqnarray}
The last terms vanish, since 
$\sum_i T_n(\mbox{\boldmath $r_{I,i}$})=T(\mbox{\boldmath $r_{I}$})$ 
for each $I$ and 
$\sum_i 
\Bigl\{2T_3(\mbox{\boldmath $r_{I,i}$})
+3T_2(\mbox{\boldmath $r_{I,i}$})-5T_1(\mbox{\boldmath $r_{I,i}$})
\Bigr\}=0$. 
Thus, we find that factors $\epsilon$ are cancelled out, 
so that the mass scale appearing in the RGE is not $M_{GUT}$ which is 
indicated by the observation but $M_*$.
It implies that the coupling unification cannot be realized 
even if we add matters in various representations.

Note that in the above argument we have assumed that all unknown
coefficients are of order unity and there are no cancellations in
diagonalizing the mass matrices. 
As is remarked in Section~\ref{sec:2}, if we allow ``tuning'' of order
$10^{-3}$ the coupling unification can be realized.

\def\Vol#1,#2,#3{{\bf #1} (#2) #3}
\def\JVol#1,#2,#3,#4{{\it #1} {\bf #2} (#3) #4}

\def\JL#1{\JVol#1}
\def\andvol#1{\Vol#1}
\def\PR#1{{\it Phys.~Rev.} \Vol#1}
\def\PRL#1{{\it Phys.~Rev.~Lett.} \Vol#1}
\def\NP#1{{\it Nucl.~Phys.} \Vol#1}
\def\PL#1{{\it Phys.~Lett.} \Vol#1}
\def\PTP#1{{\it Prog.~Theor.~Phys.} \Vol#1}

\end{document}